# Wigner Distribution of Elliptical Quantum Optical Vortex

Abir Bandyopadhyay[1, 2*] and Ravindra Pratap Singh[1†]
[1]Quantum Optics & Quantum Information Group, Theoretical Physics Division,
Physical Research Laboratory, Navrangpura, Ahmedabad – 380009, India
[2]Hooghly Engineering and Technology College, Hooghly 712103, India

## Abstract

We calculate the Wigner quasiprobability distribution function of quantum elliptical vortex in elliptical beam (EEV), produced by coupling squeezed coherent states of two modes. The coupling between the two modes is performed by using beam splitter (BS) or a dual channel directional coupler (DCDC). The quantum interference due to the coupling between the two modes promises the generation of controlled entanglement for quantum computation and quantum tomography.

## I. Introduction

Circular optical vortex beams with helical wave front can be produced in a controlled manner using methods such as computer-generated hologram (CGH), cylindrical lens mode converter, and spiral phase plate [1]. Because of their specific spatial structure and associated orbital angular momentum (OAM), they find profound applications in the field of optical manipulation [2], optical communication [3], quantum information and computation [4]. Optical vortices have drawn a great attention in the last two decades and these have prompted to start a new branch in physical optics known as singular optics [5]. However, most of the work relating to optical vortex deals with classical vortex that involves classical electromagnetic field. It is rare to find literature that takes into account of vortex formed by quantized radiation field. Agarwal *et al.* have generated quantum vortex by two mode squeezed vacuum [6]. They have shown that a two mode state under the linear transformation belonging to the SU(2) group may lead to a vortex state under special conditions [7]. However, they considered the special case of 50/50 superposition of two orthogonal fields, producing the vortex as a Fock state in circular basis [6,8].

Phase space, which is a fundamental concept in classical mechanics, remains useful when passing to quantum mechanics. On the line of probability density distribution functions in classical systems, quasiprobability distributions have been introduced in quantum mechanics. They provide a complete description of quantum systems at the level of density operators, though not at the level of state vectors. Among them, the Wigner function stands out because it is real, nonsingular, yields correct quantum-mechanical operator averages in terms of phase space integrals, and possesses positive definite marginal distributions [9]. The Wigner distribution function has come to play an ever increasing role in the description of both coherent and partially coherent beams and their passage through first order systems [10]. Once the Wigner distribution is known, the other properties of the system can be calculated from it.

In this article we propose experimental methods to prepare the generalized quantum elliptic vortex by coupling two squeezed states using beam splitter (BS) or dual channel directional coupler (DCDC). Although Agarwal et al. have studied the properties of a generic vortex wave function of the form $(x + iy)^m$, which is perfectly symmetric, however, no real physical system can exhibit perfect symmetry. Therefore, a generalized quantum elliptic vortex provides a more realistic and more widely applicable vortex model. We study the Wigner distribution of such states in detail in this article. The article is organized as follows. In section II we describe both the transformations briefly. In section III we start with the operator form of the generalized vortex, and show that they have ellipticity in the Gaussian profile as well as in the vortex structure (EEV). We then calculate and study the four dimensional Wigner quasiprobability distribution function, by projecting it in two dimension. We conclude our results in section IV.

## II. Coupling transformations (BS and DCDC)

The coupling between two beams may be performed by using beam splitter (BS) or a dual channel directional coupler (DCDC), both of which refer to SU(2) rotation. While the first one is a well utilized component in classical and quantum optics, the later one is important for quantum circuitry. Both the components are recently being used in optical coherence tomography [11]. The quantum interference due to the coupling between the two modes promises the generation of controlled entanglement for quantum computation.

*(a) Beam Splitter*

Quantum mechanically a BS couples the

annihilation operators as two input modes [8,12],

$$\begin{bmatrix} a_1(out) \\ a_2(out) \end{bmatrix} = \begin{bmatrix} A_1 & A_2 \\ A_2 & A_1^* \end{bmatrix} \begin{bmatrix} a_1(in) \\ a_2(in) \end{bmatrix} = \mathcal{U}^\dagger \begin{bmatrix} a_1(in) \\ a_2(in) \end{bmatrix} \mathcal{U}, \qquad (1)$$

where $A_i$ denotes transmitivity and reflectivity of the BS. They follow $|A_1|^2 + |A_2|^2 = 1$, and $A_1^* A_2 + A_1^* A_2 = 0$. The outputs $a_1(out)$ and $a_2(out)$ are generated by the unitary transform $\mathcal{U} = \exp\left[-i\left(\xi a_1^\dagger a_2 + \xi^* a_2^\dagger a_1\right)\right]$. The creation operators have a similar transformation, with $A_1$ conjugated and $A_2$ negated as in following Eq. (2b). The coupling generates the circular vortex $(x + iy)^m = \left(re^{i\varphi}\right)^m$ with vorticity *m*, for a 50/50 BS.

However the transmission and reflection coefficients ($A_i$) are not equal for most of the cases. In those cases the coefficients of mixing are not equal, and they produce an elliptical vortex. As the BS asymmetry becomes larger and larger, more "which path" information is available, and the quantum interference effect is correspondingly diminished. Somewhat surprisingly, this reduced interference has been found to be extremely useful in a number of quantum information processing applications in linear optics, such as, quantum computing gates [13] and quantum cloning machines [14]. Asymmetric BS has also been shown to be useful in multi-photon quantum interference experiments [15]. Because of the growing importance of asymmetric BS, it is important to fully explore the interference effects associated with them.

*(b) Dual channel directional coupler*

A lossless DCDC (see Fig. 1) couples two modes of light through evanescent waves with coupling strength $g$, dependent on the refractive indices of the two waveguides. The coupling can be controlled through Pockel's effect by applying electric field. For the coupling between two modes, under the rotating wave approximation, one finds the time evolution of the annihilation and creation operators [16, 17],

$$\begin{bmatrix} a_1(t) \\ a_2(t) \end{bmatrix} = \begin{bmatrix} A_1 & A_2 \\ A_2 & A_1^* \end{bmatrix} \begin{bmatrix} a_1(0) \\ a_2(0) \end{bmatrix}, \tag{2a}$$

and,

$$\begin{bmatrix} a_1^\dagger(t) \\ a_1^\dagger(t) \end{bmatrix} = \begin{bmatrix} A_1^* & -A_2 \\ -A_2 & A_1 \end{bmatrix} \begin{bmatrix} a_1^\dagger(0) \\ a_1^\dagger(0) \end{bmatrix}, \tag{2b}$$

where, $A_1 = \cos \Omega t - i (\delta/\Omega) \sin \Omega t$ and $A_2 = i (g/\Omega) \sin \Omega t$, with $\Omega = \sqrt{\delta^2 + g^2}$. Here we have chosen the central energy of the two oscillator modes to be zero, and defined their difference in energy as $\delta\hbar$. In case the coupling between the two modes is *not* 50/50, it results in an output in the form of the elliptical vortex. In case of the input being two different Fock states, which is product of two HG modes with different arguments in their spatial dependence, it will produce an elliptical vortex. However, with two coherent states as input the same cannot be constructed. With two squeezed states as input, one can get elliptical Gaussian distribution, by choosing the different squeezing parameters for the two modes in consideration. The action of DCDC is to produce the vorticity through coupling [7]. For a given ellipticity, i.e. $A_1/A_2$, the time for the coupling may be calculated from the knowledge of $\delta$ and $g$. The desired length of the DCDC may then be found by multiplying the desired time with velocity of light.

Our calculation does not incorporate losses in the coupling process. Losses may also be incorporated in the calculation by standard, but not so trivial master equations [8, 18], or by a specific treatment of photon-phonon interaction [12]. The lossy coupling interaction can also be modeled through a three level atom in the $\Lambda$ configuration [6, 17].

## III. Wigner distribution of generalized quantum vortex

Now, we consider two separate squeezed and displaced vacuum modes as our input states and couple them through a BS or DCDC for general case. Mixing of equal amount generates a circular vortex, which has been dealt quantum mechanically elsewhere [6,7]. The quantum mechanical description of displaced EEV (DEEV) state may be given by,

$$|\Psi_{eev}^D\rangle = \mathcal{N}\left[\eta_x a_x^\dagger \pm i\,\eta_y a_y^\dagger\right]^m S_x(\zeta_x)\,S_y(\zeta_y)\,D_x(\alpha_x)\,D_y(\alpha_y)\,|0,0\rangle, \qquad (3)$$

where $\mathcal{N}$ is the normalization constant. $S_i(\zeta_i) = exp(\zeta_i^* a_i^{\dagger 2} - \zeta_i a_i^2)$ and $D_i(\alpha_i) = exp(\alpha_i^* a_i^\dagger - \alpha_i a_i)$ are the usual squeezing and displacement operators corresponding to $x$ and $y$ directions (the index $i = x, y$). The term in square bracket, generated by BS/DCDC, is responsible for the elliptical vortex. If we put $\eta_x = \eta_y = 1$, $\zeta_x = \zeta_y = \zeta$(real) it reduces to the displaced circular vortex state in a displaced circular Gaussian beam (DCCV): $|\Psi_{ccv}^D\rangle$. $|\Psi_{ccv}\rangle$, a circular vortex in circular beam, is discussed in detail in [6] using $Q$ function. For the case $\zeta_x \neq \zeta_y$ the beam profile becomes elliptical, whereas $\eta_x \neq \eta_y$ refers to the elliptical vortex. The parameters in the generator of the vortex term $\eta_i$ are trivially connected to the reflection and transmission of the BS, as described in Eq. (1), or the coupling coefficients for DCDC, as presented in Eq. (2). The experiment in Ref [19] describes the method for implementing the effect of creation operators, which is required for producing the studied states.

Following the mathematical treatments of [6], with the choice of the parameters, $\eta_i = 1/(\sqrt{2}\,\sigma_i)$, for $i = x, y$, one can calculate the normalized spatial distribution of DEEV state as

$$\Psi_{eev}^D(x,y) = \sqrt{\frac{2^{(m-2)}}{\sigma_x \sigma_y \Gamma(m+\frac{1}{2})\sqrt{\pi}}}\left[\eta_x(x-x_0) \pm i\,\eta_y(y-y_0)\right]^m exp\left[-\frac{1}{2}\left\{\left(\frac{x-x_0}{\sigma_x}\right)^2 + \left(\frac{y-y_0}{\sigma_y}\right)^2\right\}\right], \quad (4)$$

where, $\sigma_i = exp(2\zeta_i)$. It is centered at a point $(x_0, y_0)$, where $x_0 = \mathfrak{Re}\,(\alpha_x)$ and $y_0 = \mathfrak{Re}\,(\alpha_y)$. We plot the spatial distribution $|\Psi_{eev}^D(x,y)|^2$ as a function of coordinates in Fig. (2). The plot clearly shows the elliptic structure, with zero intensity at $(x_0, y_0) = (2,4)$, the displacement of the vacuum. Inverting the ratio $\frac{\sigma_x}{\sigma_y}$ rotates the ellipse by $\pi/2$.

We change our variables to shifted (displaced) and scaled ones: $X_1 = \frac{x-x_0}{\sigma_x}$, $Y_1 = \frac{y-y_0}{\sigma_y}$, $P_{x_1} = \frac{\sigma_x}{\sqrt{2}}(p_x - p_{x_0})$, $P_{y_1} = \frac{\sigma_y}{\sqrt{2}}(p_y - p_{y_0})$, $X_2 = \frac{\sigma_y}{2\sigma_x}(x - x_0)$, $Y_2 = \frac{\sigma_x}{2\sigma_y}(y - y_0)$, $P_{x_2} = \frac{\sigma_y^3}{\sqrt{2}}(p_x - p_{x_0})$, $P_{y_2} = \frac{\sigma_x^3}{\sqrt{2}}(p_y - p_{y_0})$, with $p_{i_0} = \mathfrak{Im}(\alpha_i)$. It allows us to calculate the four dimensional Wigner function for the state $|\Psi_{eev}^D\rangle$, following the treatments of [20], in a compact fashion as,

$$W(x,y,p_x,p_y) = K\,exp\left[-\left(X_1^2 + Y_1^2 + P_{x_1}^2 + P_{y_1}^2\right)\right]L_m^{-1/2}\left[\frac{\left(P_{x_2}+P_{y_2}-X_2-Y_2\right)^2}{\sigma_x^2+\sigma_y^2}\right], \qquad (5)$$

where, $L_m^{-1/2}$ is associated Laguerre polynomial (ALP), and, $K = \frac{2^{(m-4)}\,m!}{\pi\sqrt{\pi}\Gamma(m+\frac{1}{2})}\left[-2(\sigma_x^2 + \sigma_y^2)\right]^m$. We point the fact that the effect of $D_i(\alpha_i)$ is nothing but producing a displacement of the center of the beam as well as the vortex $(x_0, y_0)$. Note that in Eq. (5) the changed variables in the Gaussian term are different from the changed variables in the argument of the ALP term. In the case of circular vortex the hole and vortex terms factor out as a product $r^{2m} L_m^0$, along with the Gaussian term. In the present case, the usual Gaussian term is factored out nicely, but the hole term $(r^{2m})$ is not separated out from the Laguerre term. We notice that it is embedded in the associated Laguerre polynomial term. One can be reminded of the Rodrigues' formula for the ALP [21] for that purpose,

$$L_m^\alpha = \frac{(-)^m}{m!} e^{x^2} x^{-2\alpha} \frac{d^m}{dx^m}\left[e^{-x^2} x^{2(m+\alpha)}\right]. \tag{6}$$

Therefore Eq. (5) ensures that the elliptical vortex may be expressed as a combination of all circular vortices from 0 to $m$.

Next we study the four dimensional Wigner function (Eq.(5)) in detail. It can be reduced to two variables by choosing the other two to be the displaced value. For such choices the Wigner function can be written as six reduced functions, each involving only two variables (2D),

$$W(x,y)_{p_y=p_{y_0}}^{p_x=p_{x_0}} = K\, exp\left[-\left(X_1^2 + Y_1^2\right)\right] L_m^{-1/2}\left[\frac{(Y_2+X_2)^2}{\sigma_x^2+\sigma_y^2}\right], \tag{7a}$$

$$W(p_x,p_y)_{y=y_0}^{x=x_0} = K\, exp\left[-\left(P_{x_1}^2 + P_{y_1}^2\right)\right] L_m^{-1/2}\left[\frac{\left(P_{y_2}+P_{x_2}\right)^2}{\sigma_x^2+\sigma_y^2}\right], \tag{7b}$$

$$W(x,p_x)_{p_y=p_{y_0}}^{y=y_0} = K\, exp\left[-\left(X_1^2 + P_{x_1}^2\right)\right] L_m^{-1/2}\left[\frac{(P_{x_2}-X_2)^2}{\sigma_x^2+\sigma_y^2}\right], \tag{7c}$$

$$W(y,p_y)_{p_x=p_{x_0}}^{x=x_0} = K\, exp\left[-\left(Y_1^2 + P_{y_1}^2\right)\right] L_m^{-1/2}\left[\frac{\left(P_{y_2}-Y_2\right)^2}{\sigma_x^2+\sigma_y^2}\right], \tag{7d}$$

$$W(x,p_y)_{y=y_0}^{p_x=p_{x_0}} = K\, exp\left[-\left(X_1^2 + P_{y_1}^2\right)\right] L_m^{-1/2}\left[\frac{\left(P_{y_2}-X_2\right)^2}{\sigma_x^2+\sigma_y^2}\right], \tag{7e}$$

$$W(y,p_x)_{p_y=p_{y_0}}^{x=x_0} = K\, exp\left[-\left(Y_1^2 + P_{x_1}^2\right)\right] L_m^{-1/2}\left[\frac{(P_{x_2}-Y_2)^2}{\sigma_x^2+\sigma_y^2}\right], \tag{7f}$$

Though all the six relations look alike up to the Gaussian terms, a careful look provides subtle differences between them. While the numerators of the arguments of the ALP terms of the first two are square of the *sum* of the two changed variables, in the other four relations the same terms are square of the *difference* of the two changed variables. Due to the presence of a square in all six such terms cross terms between the two variables appear in the Wigner functions. The first two relations provide information about the cross correlation between the same quadratures of two different modes. The relations in Eqs. (7c)-(7d) provide the information about the cross correlation between the two quadratures of the same mode. The last two relations show interesting quantum interference in the cross correlation of different quadratures of two different modes. Similar quantum interference is previously reported in [7, 12] in the same coupling transformations.

Next we show the properties of the functions in Eqs. (7a)-(7f) by plotting them in Fig (3). We choose $\sigma_x = 5$, $\sigma_y = 3$, and $m = 3$ for all the Figs. (3a-f), though we also discuss the observations for other values of the parameters. If $\sigma_i$s are interchanged, then the figures rotate by $\pi/2$. Knowing the fact that the changed variables involved in the Gaussian term is different from the changed variables in the argument of the ALP term, with the observations in the last paragraph, it is expected to observe different distribution patterns. Remembering the fact that the effect of $D_i(\alpha_i)$ is nothing but shifting the center of the beam as well as the vortex, we choose $x_0 = y_0 = p_{x_0} = p_{y_0} = 0$, without loss of any new information. In Fig. (3a) the ellipticity and vortex structure are observed for the Wigner function of two space coordinates. However, in the Fig. (3b), the Wigner function of the momentum of the different modes breaks up to two separate elliptic Gaussian functions. In the phase space of the *x* mode, Fig. (3c), new structures, beyond the core vortex, in the momentum quadratures start showing up. The number of minima matches with the value of *m* with *m*+1 maxima outside. The outermost maxima are not very clear and need to be observed carefully for the set of parameters chosen. For even *m*, it is observed that there is an even number of minima with *m*-1 maxima in between. In Fig. (3d) and (3e) the Wigner functions in the phase space of the second mode (y, $p_y$) and the cross phase space of *x* and $p_y$ show similar squeezed Gaussian structure. In Fig (3f) the dependence on y and $p_x$ shows similar dependence as in Fig. (3c). The asymmetry in the state under consideration is the reason for such similarity in the plots.

To study the quantum interference effects in the Wigner functions, we take the ratio of the cross terms, responsible for interference, and the terms containing single variable, which we call scaled interference term (SIT). The constant and the Gaussian terms cancel out, retaining only SIT in the expansion of the ALP term. This is thus the inherent property of the ALP, with a square of sum/difference of the variables involved in the argument of the ALP. For this reason, we plot them as a function of dummy indices *r* and *s*. One can expand the ALP term for different values of *m*, and study the properties of the quantum interference for various orders of the vortices. The arguments of the ALP contain square of a sum/difference of the two variables, in consideration. Noticing that if the sum is changed to difference they rotate the plot by $\pi/2$, we plot SIT in Fig. (4) only for the sum in the argument. We also plot the contours of these functions in the same figure. For $m = 1$, in Fig. (4a), we observe very sharp peaks and dips along a circle. The magnitudes of the peaks or dips are not equal, but show complicated structure, a kind of interference pattern. There are many very small structures in between the large peaks. There are peaks on the negative side also, showing the quantum nature of the states. Fig. (4b) shows the contour plot, corresponding to Fig. (4a). One may be able to notice the dots corresponding to the sharp peaks/dips in Fig (4a), if seen carefully. When we go to the higher *m*, the plots show complicated but symmetric patterns, as discussed below. For $m = 2$ in Fig. (4c) the structure breaks up into about one inner circular and another outer four lobe elliptical structures. The primary four positive peaks become higher in the outer structure, whereas the other peaks, including the negative ones, become smaller. The variations in the inner circular structure also die down. The contour plot in Fig. (4d) shows the phase (sign) of the regions clearly with the dependence of symmetry properties on the two quadratures, in consideration. In Fig. (4e), for $m = 3$, we notice that the structure breaks into one inner circular shape, contained within a four lobe elliptical shape, which is further contained in a third outer eight lobe elliptical shape. The sharp peaks of the outermost structure in the negative direction grow, while others diminish. The contour plot in Fig. (4f) does not show the different phases as in Fig. (4d). For *m* = 4, Fig. (4g-h) shows the more complicated structures around four curves. In Fig. (4g) the previously mentioned peaks in the outermost structure become sharper, while others diminish. The contour plot in Fig. (4h) shows further complicated phase structures in comparison to Fig (4d). From the contour plots, we may conclude that the odd *m* values do not show the phase variation, where as the even *m* values show this dependence.

## Conclusion

To conclude, we utilized two well known mechanisms (BS and DCDC) of coupling between two squeezed and displaced modes to generate a quantum elliptical vortex. By choosing the squeezing parameters of the two modes to be different, we embedded the elliptical vortex in an elliptical spatial distribution profile. To the best of our knowledge, previous studies on similar systems do not have any mention of the elliptical vortex, which is the main intention of this article. We have calculated the four dimensional Wigner distribution of such quantum elliptical vortex in an elliptic profile. Once the Wigner distribution is known, the other properties of the state, such as averages, variances and uncertainties, entropy and entanglement can be calculated. We notice the coupling between the different quadratures of two different modes produces quantum interferences. We have studied the interference patterns in detail and showed interesting distributions for *m* = 1- 4. The presence of interference term in the Wigner function is expected to be useful in a number of quantum information processing applications.

## Acknowledgement

AB acknowledges the Associateship at PRL. We also thank the reviewer who pointed out the Ref. [19].

___________________________________________________________________________
**\*Email: abir@hetc.ac.in; †Email: rpsingh@prl.res.in**

# References

[1] L. Allen et.al, *Progress in Optics*, **39**, ed. E. Wolf, 291 (1999); J. Leach et.al, Phys. Rev. Lett. **88**, 257901 (2002).

[2] D. G. Grier, Nature **424**, 810 (2003).

[3] G. Gibson et.al., Opt. Exp. **12**, 5448 (2004).

[4] A. Mair et.al., Nature **412**, 313 (2001).

[5] M. S. Soskin, and M. V. Vasnetsov, *Progress in optics* **41**, ed. E. Wolf, Elsevier Science, B. V., Amsterdam, 219 (2000).

[6] G. S. Agarwal, R. R. Puri, and R. P. Singh, Phys. Rev. **A56**, 4207 (1997).

[7] G. S. Agarwal and J. Banerji, J. Phys. **A39**, 11503 (2006).

[8] L. Mandel and E. Wolf, *Optical Coherence and Quantum Optics*, Cambridge Univ. Press (1995).

[9] M. Hillery et.al., Phys. Rep. **106**, 121(1984); H.-W. Lee, Phys. Rep. **259**, 147 (1995); W. P. Schleich, *Quantum Optics in Phase Space,* Wiley-VCH, Berlin, ch. 3 (2001).

[10] R. Simon and G. S. Agarwal, Opt. Lett. **25**, 1313 (2000); M. J. Bastiaans, Opt. Commun. **25,** 26 (1978); J. Opt. Soc. Am. **69,** 1710 (1979); D. Dragoman, *Progress in Optics* **37,** ed. E. Wolf, 1 (1997).

[11] A. F. Gmitro and A. R. Rouse, Optics & Photonics News **20**, 28 (2009); R. K. Wang and H. M. Subhash, *ibid*. 41 (2009).

[12] S. A. Ramakrishna, A. Bandyopadhyay and J. Rai, Opt. Exp. **2**, 29 (1998).

[13] H. Hofmann and S. Takeuchi, Phys. Rev. **A 66**, 024308 (2002); J.L. O'Brien et.al., Nature **426**, 264 (2003); K. Sanaka, et.al., Phys. Rev. Lett. **92**, 017902 (2004).

[14] R. Filip, Phys. Rev. **A 69**, 052301 (2004); Z. Zhao et.al, Phys. Rev. Lett. **95**, 030502 (2005); L. Bartuskova et.al., Phys. Rev. Lett. **99**, 120505 (2007).

[15] J. Fiurasek, Phys. Rev. A 65, 053818 (2002); H. Wang and T. Kobayashi, Phys. Rev. A 71, 021802(R) (2005); B.H. Liu et.al., Opt. Lett. 32, 1320 (2007); Phys. Rev. A 77 023815 (2008); K.J. Resch et.al. Phys. Rev. Lett. 98, 223601 (2007); Z.Y. Ou, Phys. Rev. A 77, 043829 (2008).

[16] S. Longhi, Phys. Rev. **B79**, 245108 (2009); J. Peřina, Jr., J. Peřina, in *Progress in Optics* **41**, ed. E. Wolf, 361 (2000); W. K. Lai, V. Bužek and P. L. Knight, Phys. Rev **A43,** 6323 (1991).

[17] A. Bandyopadhyay and J. Rai, Opt. Commun. **140**, 41 (1997).

[18] A. Rai, S. Das, G. S. Agarwal, "Non-Gaussian and Gaussian entanglement in coupled lossy waveguides", arxiv:0907.2432.

[19] V Parigi, A. Zavatta, Myungshik Kim, and M. Bellini, Science 317, 1890 (2007).

[20] R. P. Singh, S. Roychowdhury, V. K. Jaiswal, Opt. Commun. **274** , 281 (2007); Journ. Mod. Opt. **53** (2006) 12, 1803.

[21] I. S. Gradshteyn and R. Y. Ryzhyk, *Table of Integrals, Series, and Products*, VII Ed. (2007), Elsevier, Eq. 8.970.1.

# Figure Captions

Fig. 1. Schematic representation of a dual channel directional coupler (DCDC), shown with circular cross section, merging within a cladding material. The cross sectional dependence of the refractive index $n$ on position is shown in the right. The heights of the square peaks may vary. The propagation direction is perpendicular to the page.

Fig. 2. Plot of $|\Psi^D_{eev}(x,y)|^2$ (in arbitrary units) for $\sigma_x = 5$, $\sigma_y = 3$, $\eta_i = 1/(\sqrt{2}\,\sigma_i)$, $x_0 = 2$, $y_0 = 4$ and $m = 3$. Notice that the core of the vortex, as well as the beam is shifted to $x_0 = 2$ and $y_0 = 4$.

Fig. 3. Plot of Wigner function (in arbitrary units) (a) as a function of $x$ and $y$ (Eq. (7a)) for $m = 3$, with $\sigma_x = 5$, and $\sigma_y = 3$. (b) Same as a function of $p_x$ and $p_y$ (Eq. (7b)). (c) Same as a function of $x$ and $p_x$ (Eq. (7c)). (d) Same as a function of $y$ and $p_y$ (Eq. (7d)). (e) Same as a function of $x$ and $p_y$ (Eq. (7e)). (f) Same as a function of $p_x$ and $y$ (Eq. (7f)) .

Fig. 4. Three dimensional plots and contour plots of scaled interference terms (SIT) of Wigner function (in arbitrary units) for $m = 1$ through 4. The value of $m$ is mentioned in the subscript of SIT. (a) Surface plot of $SIT_1$ for $m = 1$. (b) Contour plot of $SIT_1$ for $m = 1$. (c-d) Surface plot and contour plot of $SIT_2$ for $m = 2$. (e-f) Surface plot and contour plot of $SIT_3$ for $m = 3$. (g-h) Surface plot and contour plot of $SIT_4$ for $m = 4$.

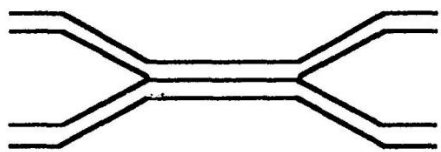
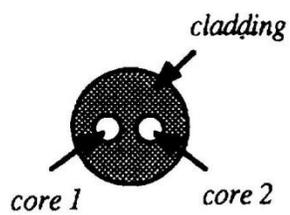
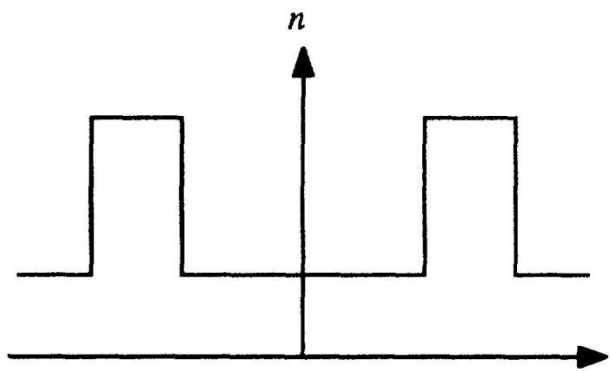

**Fig. 1.**

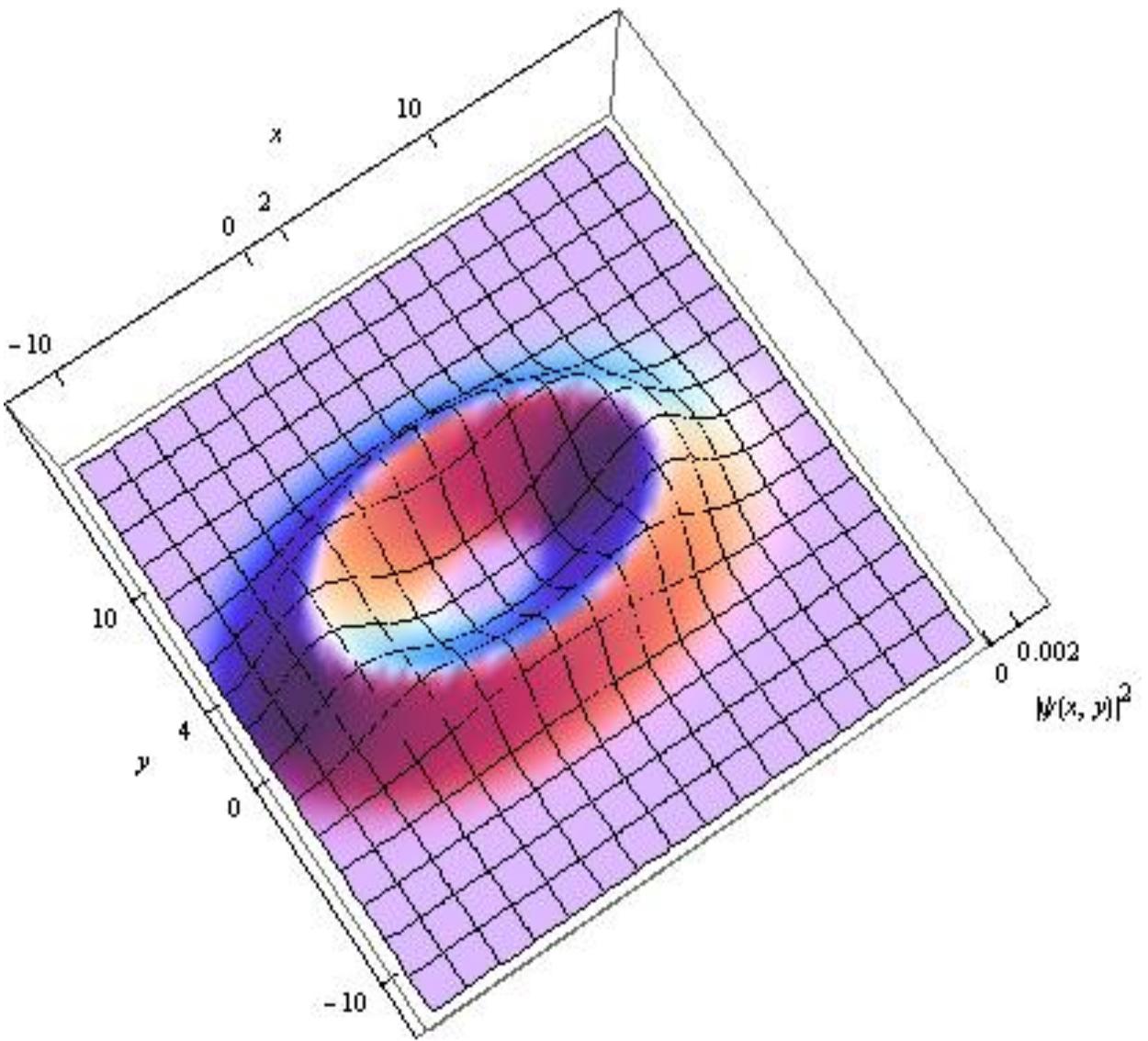

**Fig. 2.**

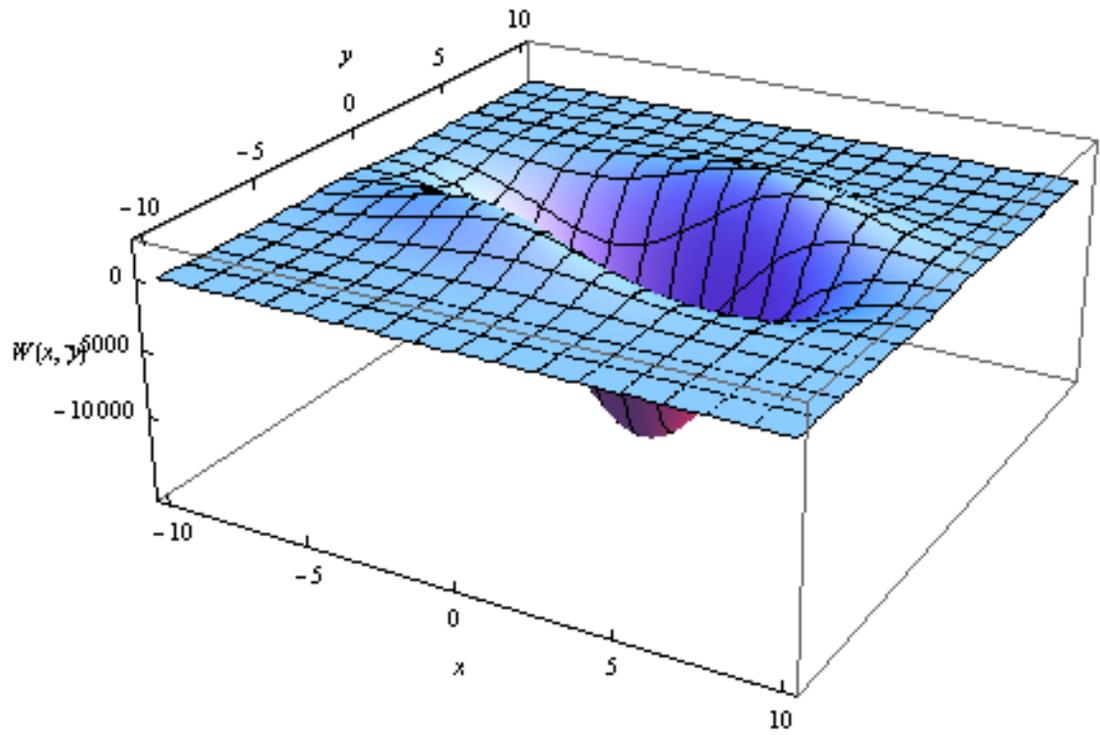

**Fig. 3a.**

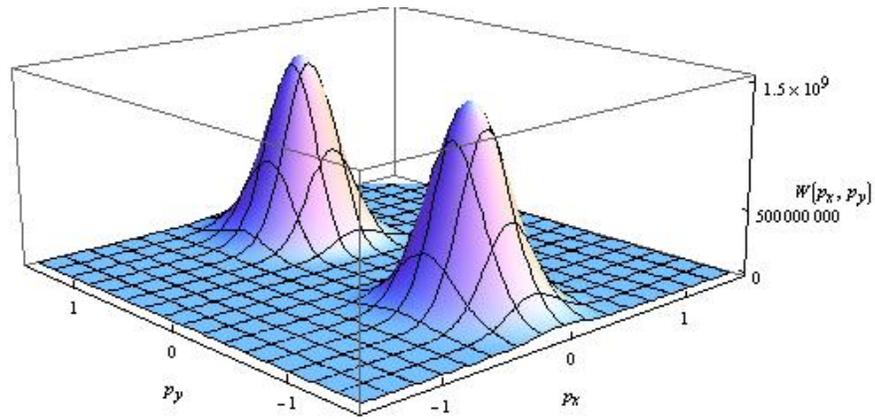

**Fig. 3b.**

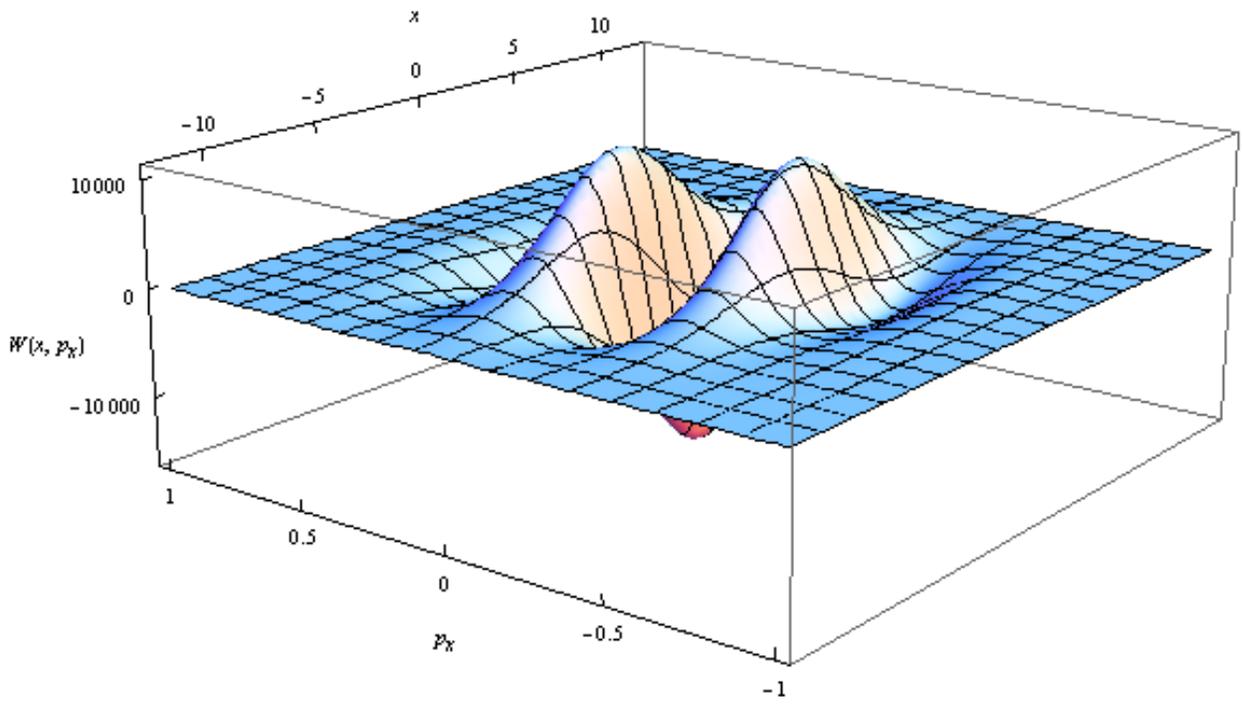

**Fig. 3c.**

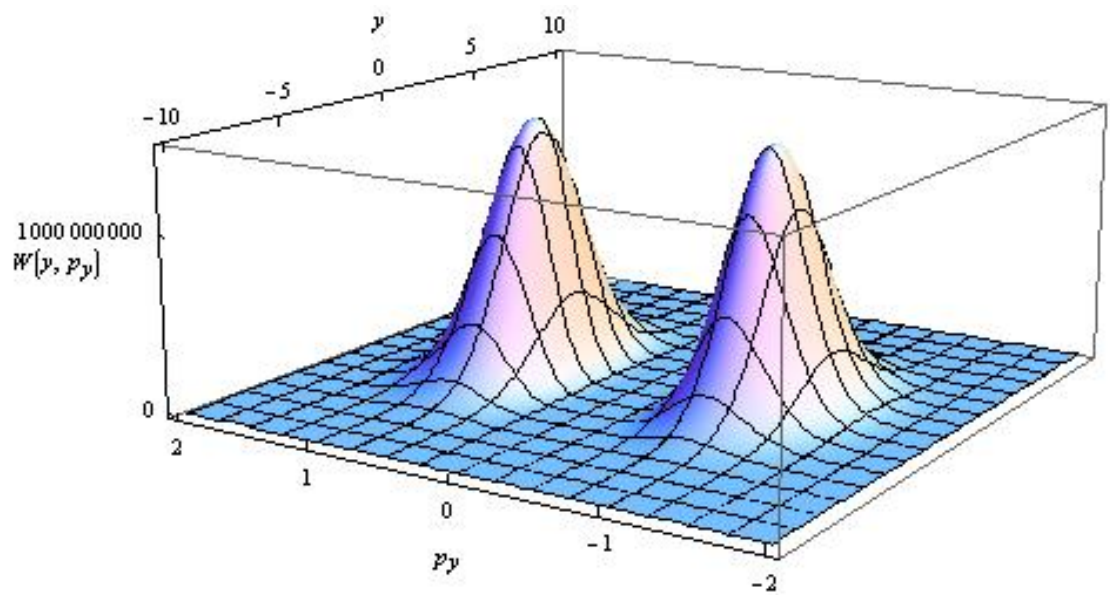

**Fig. 3d.**

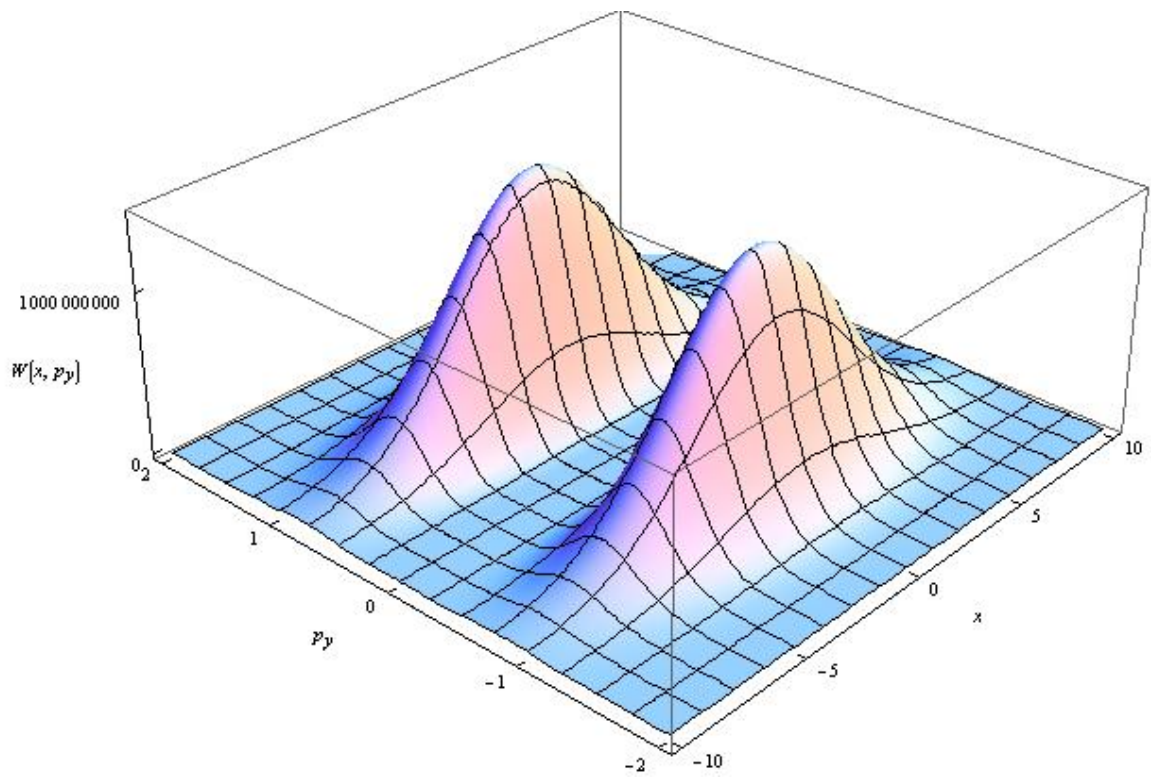

**Fig. 3e.**

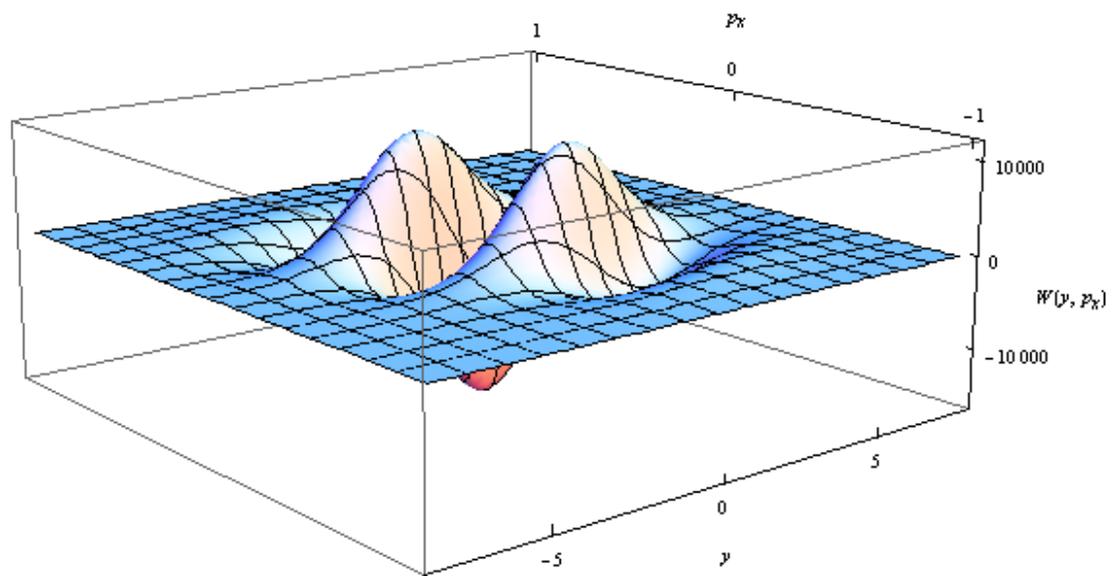

**Fig. 3f.**

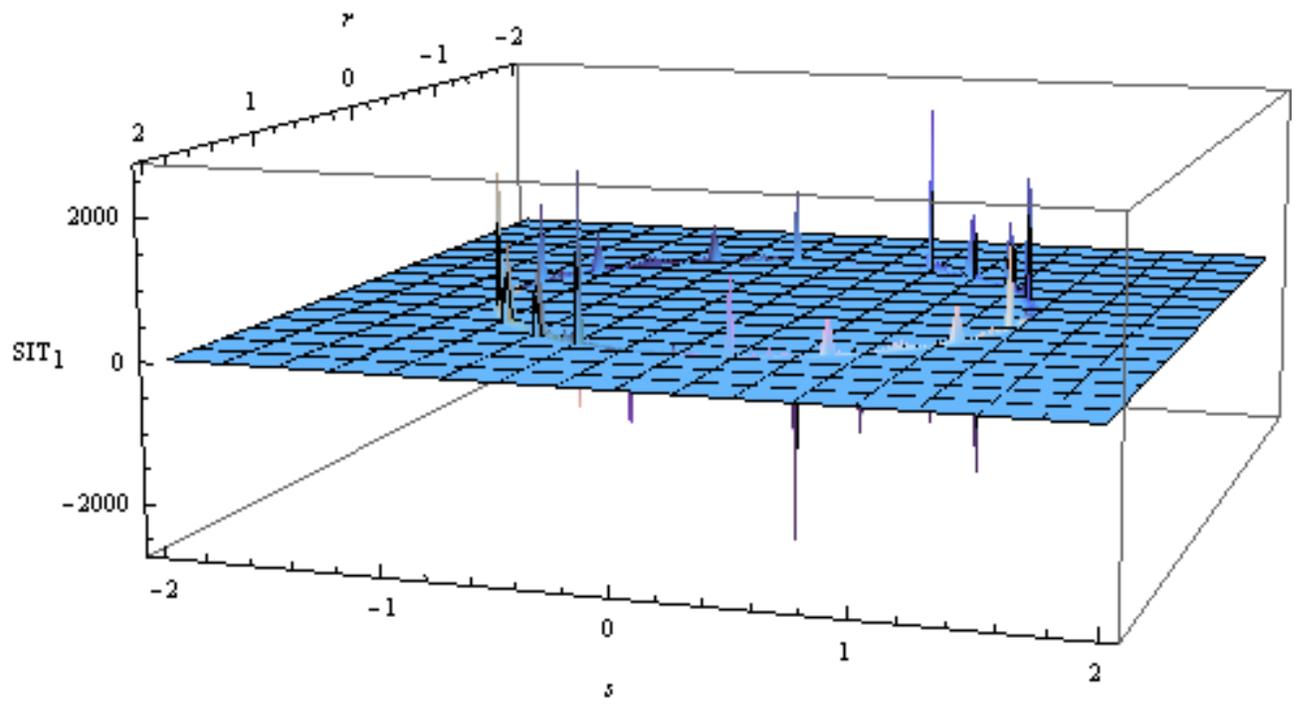

**Fig. 4a.**

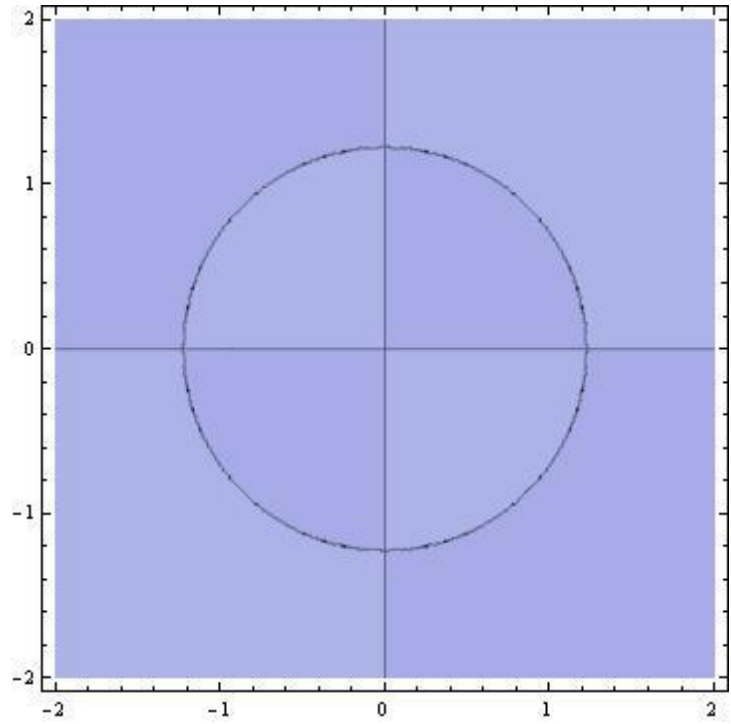

**Fig. 4b.**

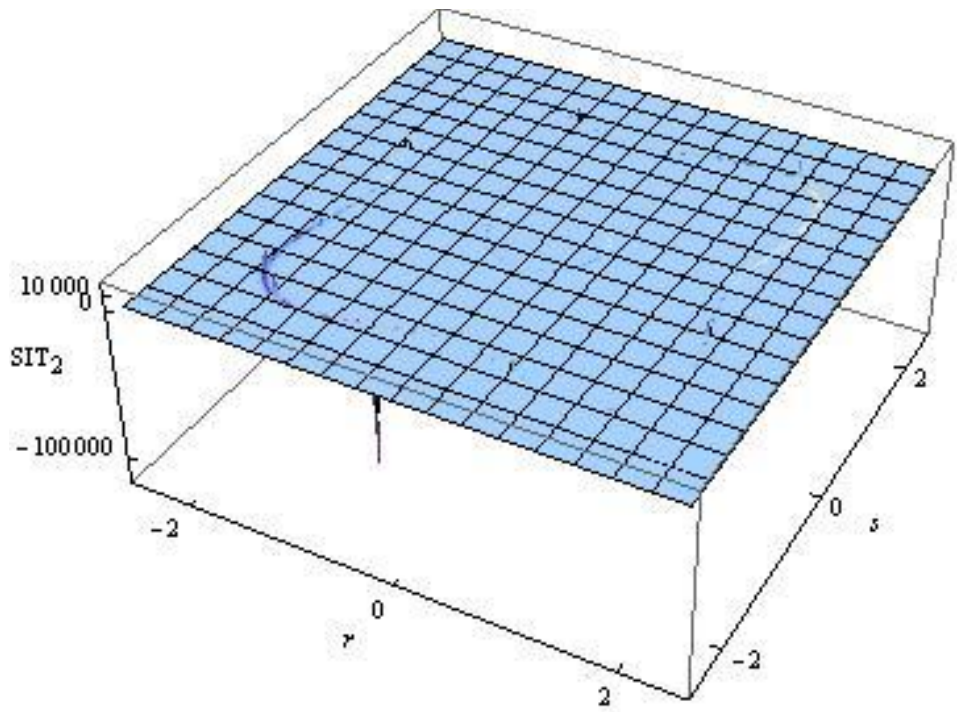

**Fig. 4c.**

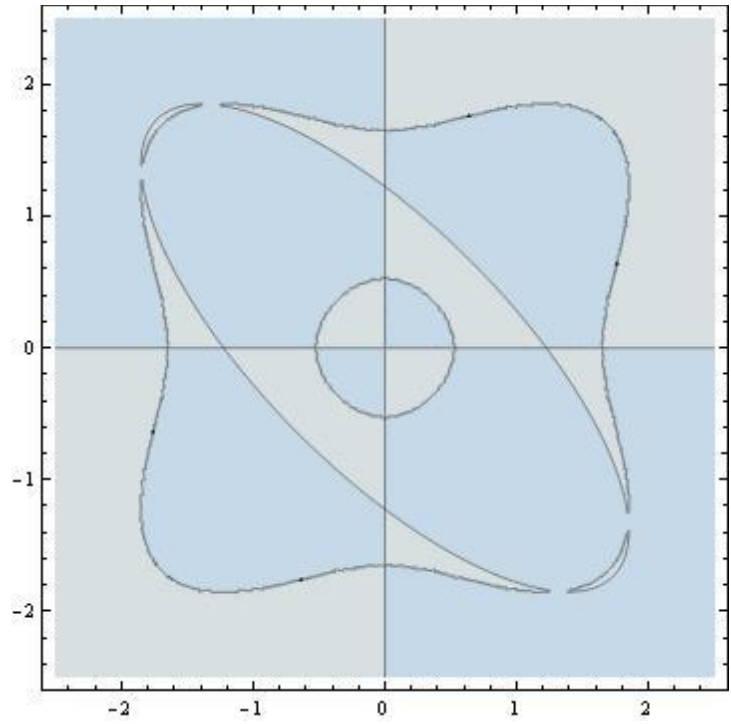

**Fig. 4d.**

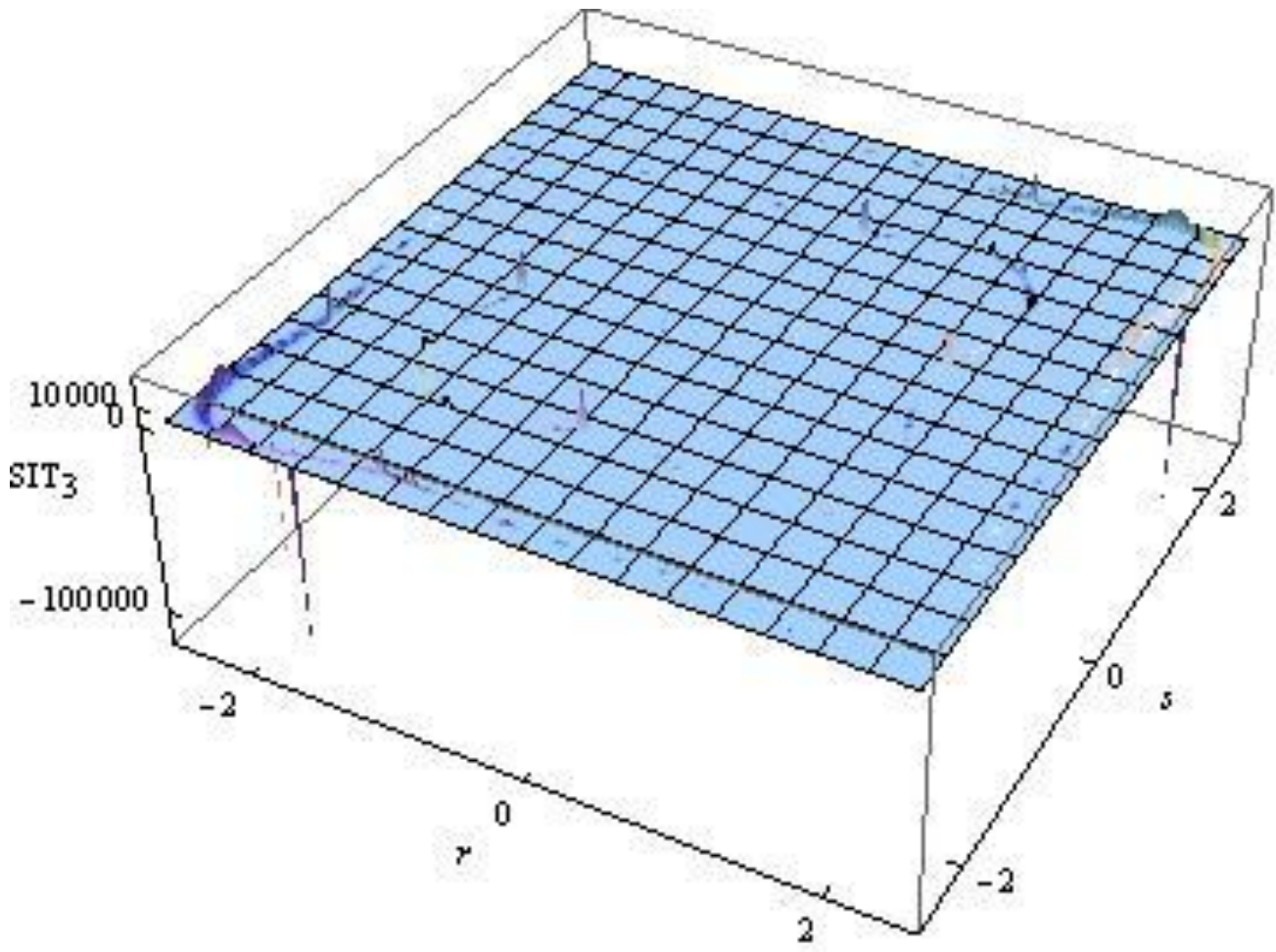

**Fig. 4e.**

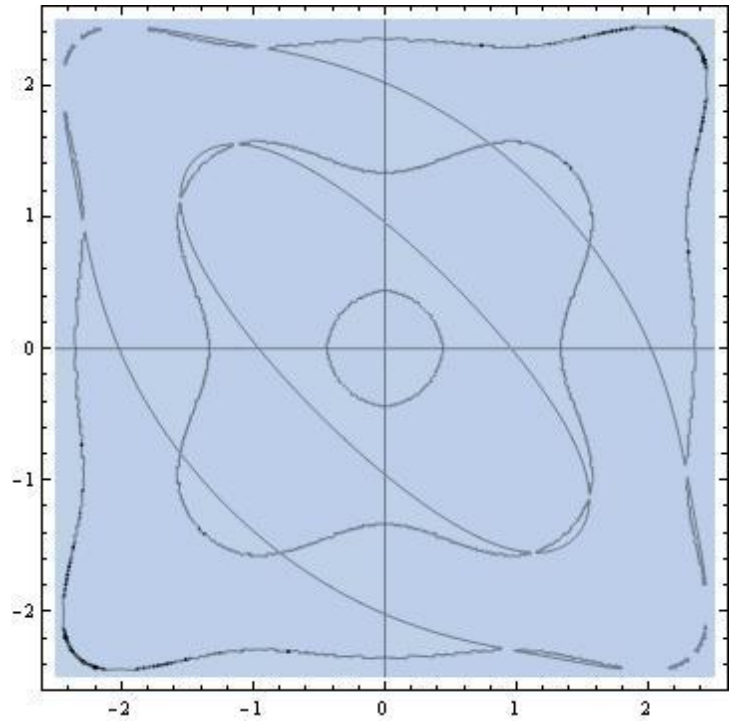

**Fig. 4f.**

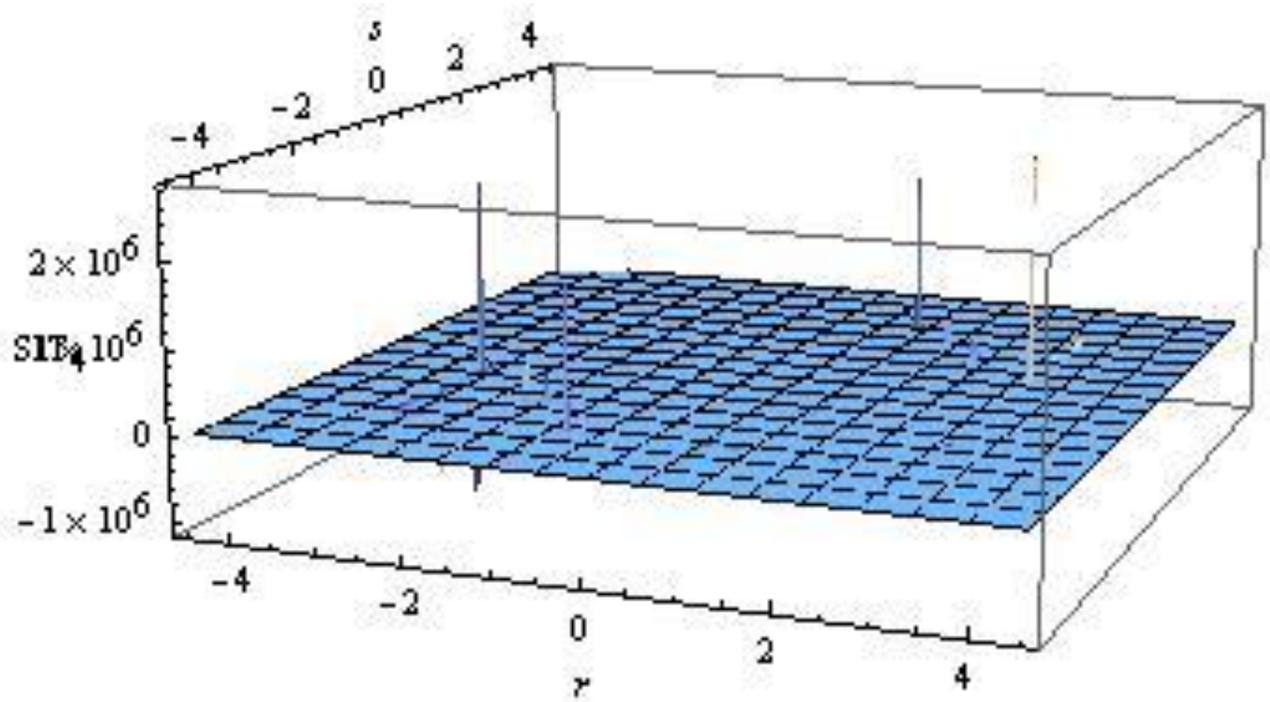

**Fig. 4g.**

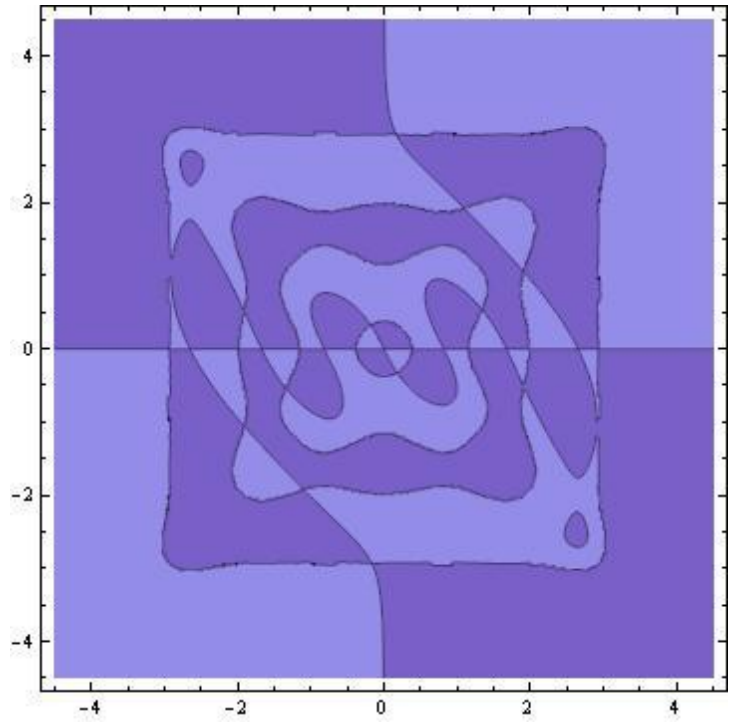

**Fig. 4h.**